\documentclass[a4paper]{article}

\usepackage{icrc2013}
\usepackage[english]{babel}
\title{FACT - Long-term Monitoring of Bright TeV-Blazars}

\shorttitle{FACT - Long-term Monitoring of Bright TeV-Blazars}

\newcommand{\uniw}{$^1$}
\newcommand{\ethz}{$^2$}
\newcommand{\tudo}{$^3$}
\newcommand{\epfl}{$^4$}
\newcommand{\unige}{$^5$}

\authors{
D.~Dorner\uniw,
A.~Biland\ethz,
T.~Bretz\ethz,
J.~Bu\ss\tudo,
S.~Einecke\tudo,
D.~Eisenacher\uniw,
D.~Hildebrand\ethz,
M.~L.~Knoetig\ethz,
T.~Kr\"ahenb\"uhl\ethz,
W.~Lustermann\ethz,
K.~Mannheim\uniw,
K.~Meier\uniw,
D.~Neise\tudo,
\mbox{A.-K.}~Overkemping\tudo,
A.~Paravac\uniw,
F.~Pauss\ethz,
W.~Rhode\tudo,
M.~Ribordy\epfl,
T.~Steinbring\uniw,
F.~Temme\tudo,
J.~Thaele\tudo,
P.~Vogler\ethz,
R.~Walter\unige,
Q.~Weitzel\ethz,
M.~Z\"anglein\uniw $\;\;$
(FACT Collaboration)
}
\afiliations{
\ethz ETH Zurich, Switzerland $\;\;-\;\;$
   Institute for Particle Physics, Schafmattstr.~20, 8093 Zurich\\
\tudo Technische Universit\"at Dortmund, Germany $\;\;-\;\;$
   Experimental Physics 5, Otto-Hahn-Str.~4, 44221 Dortmund\\
\uniw Universit\"at W\"urzburg, Germany $\;\;-\;\;$
   Institute for Theoretical Physics and Astrophysics,
   Emil-Fischer-Str.~31, 97074 W\"urzburg,\\
\epfl EPF Lausanne, Switzerland  $\;\;-\;\;$
   Laboratory for High Energy Physics, 1015 Lausanne\\
\unige University of Geneva, Switzerland $\;\;-\;\;$
   ISDC Data Center for Astrophysics, Chemin d'Ecogia~16, 1290 Versoix \\
}

\email{dorner@astro.uni-wuerzburg.de}

\abstract{Since October 2011, the First G-APD Cherenkov Telescope
(FACT) is operated successfully on the Canary Island of La Palma. Apart
from the proof of principle for the use of G-APDs in Cherenkov
telescopes, the major goal of the project is the dedicated long-term
monitoring of a small sample of bright TeV blazars. The unique
properties of G-APDs permit stable observations also during strong moon
light. Thus a superior sampling density is provided on time scales at
which the blazar variability amplitudes are expected to be largest, as
exemplified by the spectacular variations of Mrk\,501 observed in June
2012.\\ While still in commissioning, FACT monitored bright blazars
like Mrk\,421 and Mrk\,501 during the past 1.5 years so far.\\
Preliminary results including the Mrk\,501 flare from June 2012 will be
presented.}


\begin{document}
\maketitle

%Begin a section.
\section{Introduction}

The main goal of the First G-APD Cherenkov Telescope (FACT), apart from
the proof of principle for the solid state photo sensors, is the long
term monitoring of bright TeV Blazars \cite{bib:design}. 

Providing a stable and robust detector
\cite{bib:feedback,bib:predictions}, G-APDs allow for observations
during strong moon light \cite{bib:moon} enlarging the duty cycle of
the telescope which provides a better sampling of the long-term light
curves. Having a stable detector, a robust quick look analysis can be
established allowing to send flare-alerts to other telescopes within a
short time range. This allows to study these flares with more sensitive
instruments and in different wavelengths. 

However the main physics goal of FACT is to study TeV blazars not only during
flares, but in all flux states allowing for statistical analyses of the
variability of these objects on an unbiased data sample. 

Together with long-term monitoring in the optical and radio band,
detailed multi-wavelength (MWL) studies can be carried out. 

%missing: physics motivation

\section{Observations}

In January 2012, the  monitoring of bright TeV blazars with FACT was
started. Apart from few other sources, in the first 1.5 years mainly
the Crab Nebula and the two active galactic nuclei (AGN) Mrk\,421 and
Mrk\,501 have been observed. All monitoring observations were carried
out in wobble mode, i.e.\ the source is tracked 0.6 degree from the
camera center alternating for different positions. The two wobble
positions were chosen such that bright stars are either not in the FoV
or, in case it could not be avoided, are symmetric to the source for
the two positions. 

%Data with bad weather and technical problem were rejected from the data
%set. Details on the data check and the selected sample will be provided.
%%in the final version of the proceedings. 

%data sample, data check

%Crab, Mrk\,501, Mrk\,421, more sources? 

\section{Data Analysis}

The data analysis consists of several steps and is done with the
software package 'MARS - CheObs ed.' \cite{bib:mars}. First, the
data are calibrated and the signal is extracted. After the so-called
image cleaning, i.e.\ removing pixels which do not contain any signal,
the images, parameters describing the shower images are calculated.
Based on these parameters, the background is suppressed.  With the
parameter theta, which describes the angle between the reconstructed
shower origin and the real source position, the excess is determined by
subtracting the background from the signal events in the source region.
Dividing the number of excess events by the effective ontime of the
observations, excess rates are obtained. For a real flux calculation,
the engergy of the primary particle of each shower has to be
reconstructed which needs Monte Carlo simulations. Work on these
siulations and the determination of the energy spectrum are ongoing. 

The analysis chain, providing the excess rate curves, runs both on site
in La Palma and at the data center with only minor differences. In the
data center, all data are reprocessed once a newer software version is
available, while the quick look analysis (QLA) does not include
reprocessing, but just uses the latest stable and well tested software
version. Furthermore, the QLA does not include any data check so far. 

The aim of the QLA is to provide results as fast as possible to allow
for sending flare alerts to other telescopes during the same night.
Within 30 minutes to 2 hours after the data were taken, the results are
available on a dedicated webpage. To achieve this fast response on
site, this analysis does not use Monte Carlo simulations. 

Once the data are transferred to the data center, the analysis is
repeated and refined. 

Based on the measured performance of the detector
\cite{bib:feedback,bib:predictions,bib:moon}, a detailed data check is
done, and the good quality data are further analysed. 

To allow for a non-biased excess rate curve, the dependence of the
excess rate on the ambient light and on the zenith distance is studied
using about 350 hours of data from the Crab Nebula. The dependence on
the zenith distance is studied using a data sample taken during dark
time only. To disentangle the two effects, the dependence on ambient
light is studied with data with small zenith distance. Fitting the
dependencies, an estimate of the flux corrected for the influence of
ambient light and zenith distance can be calculated.

%description of analysis chain, explanation of correction of threshold
%for effect of currents and zd

\section{Data Selection}

For the presented excess rate curves, only data with well-understood,
stable performance of the detector have been used. Data taken before
the current control current control was implemented in the feedback
system \cite{bib:feedback} in April 2012 is not included. Data between
taken 6.12.2012 and 10.1.2013 were excluded because of a broken bias
voltage channel. These data may be recovered once the telescope
performance of these time periods is studied in more detail. 

\begin{figure}%[t]
\begin{center}                
 \includegraphics*[width=0.4\textwidth]{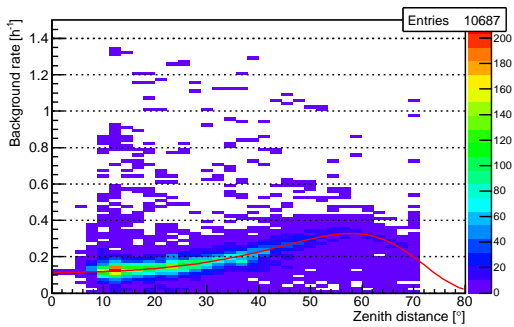}
\caption{Background rate after cuts versus zenith distance. The colour 
code represents the number of runs, where one run comprises five
minutes.  For the study, almost 900 hours of data have been used.
Details on how this background rate is determined can be found in
section 4.} 
\label{fig:bgrate} 
\end{center}
\end{figure}

To exclude data with bad weather the background rate after cuts has
been studied. With this, also data affected by laser shots from
neighbouring atmosheric moinitoring devices can be excluded. 

For this the dependency of the rates are studied versus zenith distance
(see figure \ref{fig:bgrate}) and threshold for a data set of almoust 900
hours from a time range between May 2012 and June 2013 also excluding
the data with different telescope setup mentioned above. 

For this study, the background rate after background suppression cuts,
but before theta-cut is used. To have no bias e.g.\ from strong flares,
the number of signal events is subtracted before calculating this rate.

By fitting event distributions of the obtained background rate in
zenith bins, the mean rate per zenith distance bin has been determined.
These mean values are plotted versus zenith distance and fitted. With
the determined fit function (see red line in figure \ref{fig:bgrate}),
the background rate can be corrected for the dependency on zenith
distance.

The same procedure is applied for the dependency on the trigger
threhold yielding a value of the background rate which is independent
of these two observation conditions. In the resulting event
distribution, good data can be identified. Data with a corrected
background rate deviating from these typical values are rejected, where
data with a low corrected background rate could be identified as
affected by bad weather and data with a very high corrected
background rate as affected by laser shots from a neighbouring
atmospheric monitoring device (LIDAR \cite{bib:lidar}) and is rejected
as well.  This data selection is done on run-basis, where one run lasts
five minutes. 

For the nightly excess rate curves, also nights with less than 20
minutes remaining observation time have been rejected. 

\section{Results}

\begin{figure*}%[t]
\begin{center}                
 \includegraphics[width=\textwidth]{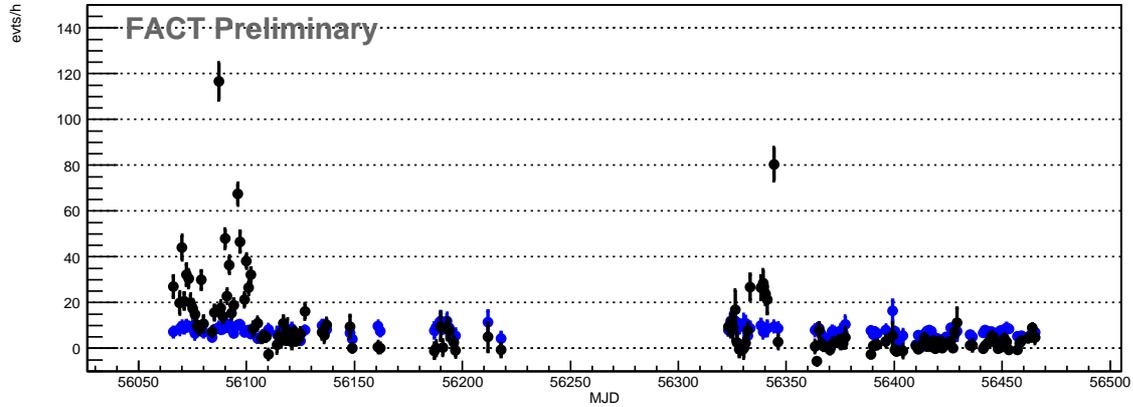}
\caption{Rates for Mrk\,501 from May 2012 until June 2013 with one bin
per night. In black: Rate of excess events. In blue: Rate of background
events. The curves are not yet corrected for the effects of the zenith
distance and of ambient light like the moon. The gap in the middle of 
the curve is due to the source being below the horizon or only visible
at very big zenith distance. In June 2012 a big outburst can be seen,
see in detail figure \ref{fig:mrk501flare}, and another one in February
2013. } 
\label{fig:mrk501} 
\end{center}
\end{figure*}

\begin{figure*}%[!t]
\begin{center}                
 \includegraphics[width=\textwidth]{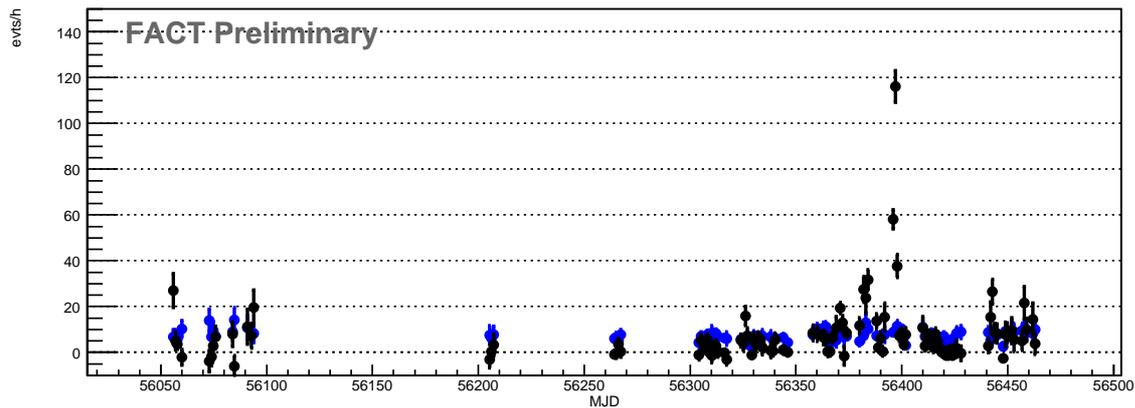}
\caption{Rates for Mrk\,421 from May 2012 until June 2013 with one bin
per night. In black: Rate of excess events. In blue: Rate of background
events. The curves are not corrected for the effects of the zenith
distance and of ambient light like the moon. The gap after the first 
data points is due to the source being not visible for a few  months
under zenith distances smaller than 60 degree. The data points in the
middle of the gap are taken under very large zenith distance due to an
historical outburst being reported in the radio range. In winter
2012/2013 there is another gap of few weeks due to a broken bias
channel. In April 2013, a major outburst is visible.} 
\label{fig:mrk421}
\end{center} 
\end{figure*}

Since the start of operation in October 2011, FACT monitored Mrk\,421
already from 25.1.2012 until now. For Mrk\,501, the observations
started on 19.5.2012, and the monitoring of both sources is ongoing.
From the QLA on site, the excess rate curves are determined as
described in section 3 and shown in the figures \ref{fig:mrk421} and
\ref{fig:mrk501} after data selection as described in section 4. Along
with the excess rates (blue), the background rates (black) are shown,
both with daily binning. It has to be noted that the plots shown here
are not yet taking into account any corrections for the effect of
ambient light and zenith distance. This explains some systematic
changes of the background over time. 

The work on data check and the determining the correction factor based
on the above explained study on Crab data are ongoing. Nevertheless,
these corrections are small compared to the large fluctuations seen
due to major flaring activities of both AGNs. 

\begin{figure}%[t]
\begin{center}                
 \includegraphics*[width=0.48\textwidth,angle=0,clip,trim=0 0 1.7cm 1.3cm]{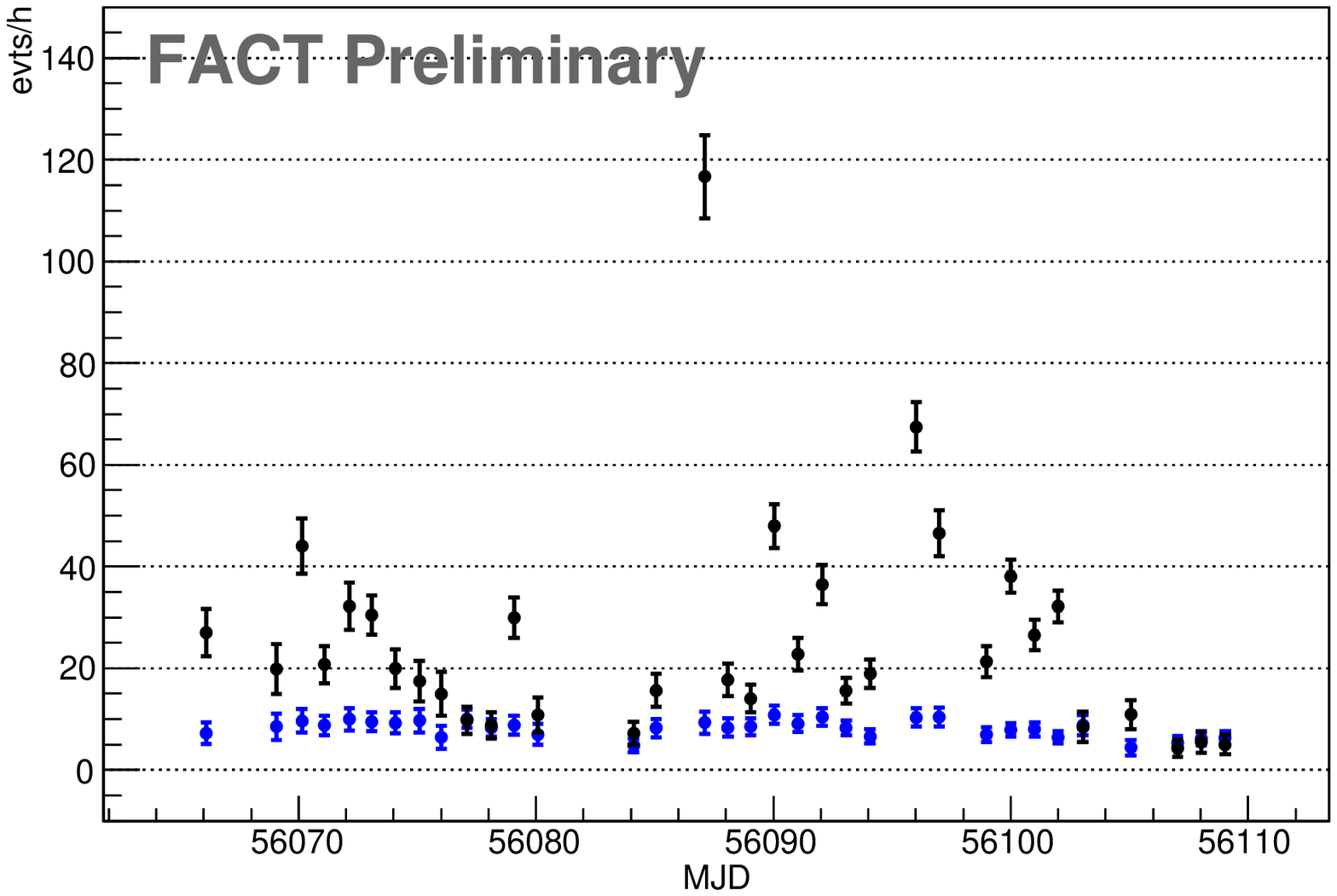}
\caption{Rates for Mrk\,501 from May 2012 until June 2012 with one bin
per night. In black: Rate of excess events. In blue: Rate of background
events. The curves are not yet corrected for the effects of the zenith
distance and of ambient light like the moon. This zoom on the excess 
rate curve shows only the time around the flare in June 2012. } 
\label{fig:mrk501flare} 
\end{center}
\end{figure}

From Mrk\,501, a major outburst could be detected in June 2012 during a
MWL campaign. During the night with the highest flux, FACT measured a
signal from the source with more than 5\,sigma significance in
5~minutes over the whole night. In figure \ref{fig:mrk501flare}, a
nightly excess rate curve for the time between 19.5.2012 and 30.6.2012
is showning the flare activity in more detail. In February 2013,
another major flare of Mrk501 was observed. 

The excess rate curve of Mrk421 is shown in figure \ref{fig:mrk501}. In
April 2013, a major outburst from the source was observed where also a
wide coverage of observation time with other telescopes is available.
However, FACT did not observe during the two nights with the highest
flux due to technical problems with the drive system because of too
high temperatures in the container. In addition, it was known that the
big telescopes are observing the source. 

The excess rate curves spanning more than one year for both sources
show nicely the monitoring power of the instrument which provides
useful information both for MWL studies as well as for variability
studies of bright TeV blazars. Due to the fast response of the QLA,
FACT is now able to trigger observations of other instruments in case
of a major outburst. 

%In the final version of the proceedings, excess rate curves will be
%shown. 

%Excess rate curves for the monitored TeV Blazars will be shown. 

%corrected excess rates (daily binning), signal of complete sample,
%observed flares, zoom on flares, 5 sigma in 5 min

\section{Conclusions and Outlook}

G-APDs have shown to be ideal for monitoring, as they allow for
observations during strong moon light. They provide stable and robust
performance of the detector. Furthermore they show to be a good
alternative for cheap monitoring telescopes which can be operated
remotely and automatic. Based on the experiences gained so far with
FACT, the long-term objective is to build several small telescopes and
realize 24/7 monitoring of bright TeV Blazars.

Since its first light, FACT has collected 1.5 years of monitoring data
at TeV energies, minimizing the gaps due to full moon and providing a
data sample which can be used both for flare and MWL studies.  During
that time, FACT has seen three major outburst of TeV blazars, one of
Mrk421 and two of Mrk501. 

With its quick look analysis, FACT furthermore can now provide fast
alerts to other telescopes in case of flares. The results of the QLA
will be available on \mbox{http://www.fact-project.org/monitoring}
starting from September 2013. 

\vspace*{0.5cm}

\footnotesize{{\bf Acknowledgement:}{The important contributions from
ETH Zurich grants ETH-10.08-2 and ETH-27.12-1 as well as the funding by
the German BMBF (Verbundforschung Astro- und Astroteilchenphysik) are
gratefully acknowledged. We are thankful for the very valuable
contributions from E.\ Lorenz, D.\ Renker and G.\ Viertel during the
early phase of the project We thank the Instituto de Astrofisica de
Canarias allowing us to operate the telescope at the Observatorio Roque
de los Muchachos in La Palma, and the Max-Planck-Institut f\"ur Physik
for providing us with the mount of the former HEGRA CT\,3 telescope,
and the MAGIC collaboration for their support. We also thank the group
of Marinella Tose from the College of Engineering and Technology at
Western Mindanao State University, Philippines, for providing us with
the scheduling web-interface. }}

\end{document}